\documentclass{article}
\usepackage{spconf}
\usepackage[utf8]{inputenc}
\usepackage[T1]{fontenc}
\usepackage[ruled,vlined]{algorithm2e}
\usepackage{algorithmicx}
\usepackage{booktabs}
\usepackage{array}
\usepackage{tabularx}
\usepackage{amsmath}
\usepackage{graphicx}
\usepackage{multirow}
\usepackage{longtable}
\usepackage{amssymb}

\def\ie{{\it i.e., }}   
\def\etc{{\it etc }}

\def\R{\mathbb R}  
\title{HIDDEN MARKOV RANDOM FIELDS AND CUCKOO SEARCH METHOD FOR MEDICAL IMAGE SEGMENTATION}
\name{EL-Hachemi Guerrout$^{1}$, Ramdane Mahiou$^{1}$, Dominique Michelucci$^{2}$, Boukabene Randa$^{1}$ and Ouali Assia$^{1}$}
\address{$^{1}$Ecole nationale Sup\'erieure en Informatique,
		Laboratoire LMCS,
		Oued-Smar, Algiers, Algeria\\
		\{e\_guerrout, r\_mahiou\, b\_randa, o\_assia\}@esi.dz}
\address{$^{1}$Ecole nationale Sup\'erieure en Informatique,
	Laboratoire LMCS,
	Oued-Smar, Algiers, Algeria\\
	\{e\_guerrout, r\_mahiou\, b\_randa, o\_assia\}@esi.dz \\	
	$^{2}$Dominique Michelucci
Universit\'e de Bourgogne,
Laboratoire LE2I,
Dijon, France\\
dominique.michelucci@u-bourgogne.fr
}

\begin{document}
	%
\maketitle
\begin{abstract}
Segmentation of medical images is an essential part in the process of diagnostics. Physicians require an automatic, robust and valid results. Hidden Markov Random Fields (HMRF) provide powerful model. This latter models the segmentation problem as the minimization of an energy function. Cuckoo search (CS) algorithm is one of the recent nature-inspired meta-heuristic algorithms. It has shown its efficiency in many engineering optimization problems. In this paper, we use three cuckoo search algorithm to achieve medical image segmentation. 

\end{abstract}


\section{INTRODUCTION}
With the overwhelming number of medical images, the manual analysis and  interpretation of images from different imaging modalities
(Radiography, Magnetic Resonance Imaging (MRI), Computed Tomography (CT),  \etc) became a tedious task. This fact underlines the necessity of automatic image  analysis, through several operations including segmentation.

Segmentation is the process of partitioning an image into multiple segments semantically interpretable. For MRI, segmentation is hard because MR images are usually corrupted by noise and non-uniformity artifact due to various factors, such as partial variations in illumination or radio frequency coil in image acquisition devices. 

In this work, we use Hidden Markov random field (HMRF) model \cite{1984-geman-geman,deng2004unsupervised,zhang2001segmentation} for image segmentation:  the  hidden information, namely the segmented image, is computed from the original image by maximizing the MAP criterion  (Maximum  A Posterior). Since the objective function of the MAP is non-linear, complex and may have several local maxima,  meta-heuristic algorithms are preferred for  optimization.

The use of meta-heuristic algorithms for image segmentation is not new. Several methods have been proposed in the literature, for instance, the cuckoo search (CS) algorithm. It is one of the latest nature-inspired meta-heuristic algorithms, it was proposed in 2009 by Xin-She Yang and Suash Deb \cite{yang2009cuckoo}. CS is based on the brood parasitism of some cuckoo species. In addition, this algorithm is enhanced by the so-called Lévy flights rather than by simple isotropic random walks.

In this paper, we compare three variants of cuckoo search (standard CS \cite{yang2009cuckoo,yang2010engineering,yang2010nature}, improved CS \cite{valian2011improved} and auto adaptative modified CS \cite{li2015modified}) to optimize the HMRF model for MRI  segmentation. In this specific  case, the segmentation consists in partitioning the brain image into different parts that are: gray matter(GM), white matter (WM) and cerebrospinal fluid (CSF). The quality evaluation is conducted on ground-truth images from BrainWeb and IBSR databases, using the Dice coefficient metric \cite{1945-dice}.  

The rest of the paper is organized as follows. In Part II, the HMRF model principles in the context of image segmentation is described. The combination of HMRF and cuckoo search algorithm is explained in Part III. Finally, in Part IV and Part V, experimental results on medical image samples are illustrated and conclusions are drawn, respectively.

\section{HIDDEN MARKOV RANDOM FIELD MODEL}
Let $y=({y}_{1},{y}_{2},{\dots},{y}_{M })$ be the image to segment into $K$ classes and $x=({x}_{1},{x}_{2},{\dots},{x}_{M })$ be the segmented image. $y_s$ is pixel value of the site $s$ that takes its values in the gray level space $ E_{y}=\{0,\ldots,255\} $. $x_s$ is class of the site $s$ and takes its values in the discrete space $E_x=\left\{1,{\dots},K\right\}$.

The image to segment $y$ and the segmented image $x$ represent respectively realizations of Markov Random Fields $Y=({Y}_{1},{Y}_{2},{\dots},{Y}_{M })$ and $X=({X}_{1},{X}_{2},{\dots},{X}_{M })$.
Configurations set of the image to segment $y$ and of the segmented image are respectively $\Omega_y=E_{y}^M$ and $\Omega_x=E_{x}^M$. 

	Let $\mu=(\mu_1,\dots,\mu_j,\dots,\mu_K)$ be the means and $\sigma=(\sigma_1,\dots,\sigma_j,\dots,\sigma_K)$ be the standard deviations of the $K$ classes in the segmented image $x=(x_1,\dots,x_s,\dots,x_M)$ \ie
\begin{equation}{\label{sj}}
\begin{array}{l}
\begin{cases}
\mu_j={\frac{1}{|S_j|} \sum_{s \in S_j} y_s}\\
\sigma_j=\sqrt{\frac{1}{|S_j|} \sum_{s \in S_j} (y_s-\mu_j)^2}\\
S_j=\{s\ |\ x_s=j\}
\end{cases}
\end{array}
\end{equation}

Hidden Markov Random Field approach is modelled in \cite{guerrout2018conjugate} as a minimization of the function $\Psi(\mu)$ defined below. We can always compute $x$ through $\mu$ by classifying $y_s$ into the nearest mean $\mu_j$ \ie $x_s=j$ if the nearest mean to $y_s$ is $\mu_j$. Thus, we look for $\mu^*$. The configuration set of $\mu$ is $\Omega_{\mu}=[0\dots255]^K$.
\begin{equation}\label{e-psi-mu}
\begin{array}{l}
\begin{cases}
{\mu}^*=\operatorname*{arg}_{\mu \in \Omega_{\mu}}{\mathit{\min}}\left\{\Psi(\mu)\right\}\\\\
\Psi(\mu) = \sum_{ j = 1 }^{K} f(\mu_j)\\\\
f(\mu_j) = \sum\limits_{s \in S_j }{ [\ln (\sigma_j ) + \frac{ (y_s - \mu_j)^2 }{ 2 \sigma_j^2 }] } \\ \quad \quad \quad \quad +  \frac{B}{T} \sum_{c_2 = \{s,t\}}{ ( 1 - 2 \delta(x_s,x_t) ) } 
\end{cases}
\end{array}
\end{equation}
\noindent where $B$ is a constant, $ T $ is a control parameter called temperature,   
$ \delta $ is Kronecker's delta and $S_j$, $\mu_j$ and $\sigma_j$ are defined in (\ref{sj}).
When  $B>0$, the most likely segmentation corresponds to the constitution of large homogeneous regions. The size of these regions is controlled by the parameter $B$.

To apply unconstrained optimization techniques, we redefine the function $\Psi(\mu)$ for $\mu \in \R^K$ instead of $\mu \in [0 \dots 255]^K$ as recommended by \cite{boyd2004convex}. Therefore, the new function  $\Psi(\mu)$ becomes as follows:

\begin{equation}\label{psi2}
\Psi(\mu) =\sum_{ j = 1 }^{K} F(\mu_j)\ \ \mbox{where   } \mu_j \in \R 
\end{equation}

\begin{equation*}
F(\mu_j)=
\begin{cases}
f(0)-u_j*10^3 & \mbox{if } \mu_j < 0\\
f(\mu_j) & \mbox{if }  \mu_j \in [0\dots255]\\
f(255)+(u_j-255)*10^3 & \mbox{if } \mu_j > 255
\end{cases}
\end{equation*}

\section{Hidden Markov Random Field and Cuckoo Search (CS) algorithm}
 To solve the minimization problem expressed in the \ref{psi2}, we used Cuckoo Search (CS) algorithm. The principal steps are set out below.

 Each egg in a host nest represents a solution $\mu^{i,t}=(\mu^{i,t}_1,\dots,\mu^{i,t}_j,\dots,\mu^{i,t}_K)$ at time $t$.
 
 Let $c^{i,t}=(c^{i,t}_1,,\dots,c^{i,t}_j,\dots,c^{i,t}_K)$ be cuckoo egg generated at time $t$. 
 
 Let $n$ be the number of available host nests (or different solutions). The initial population $\{\mu^{i,0}\}_{i=1,\dots,n}$ is generated by random initialization. 
 
 Let $\mbox{best}^t=(\mbox{best}^t_1,\dots,\mbox{best}^t_j,\dots,\mbox{best}^t_K)$ be the best solution at time $t$. 
 \begin{equation}\label{best}
  \mbox{best}^t:=\arg\min\limits_{\mu^{i,t}_{i=0,\dots,n} } {\Psi(\mu^{i,t})} 
 \end{equation}
 
 Cuckoo Search (CS) algorithm is based on three rules defined in \cite{yang2009cuckoo,yang2010engineering} and proceeds as follows:

\begin{enumerate}
	\item Generating new cuckoos $\{c^{i,t}\}_{i=1,\dots,n}$ can be performed as follows (L\'evy flight):
	\begin{equation}\label{cuckoos}
	c^{i,t}:=\mu^{i,t}+ \alpha \times \mbox{\mbox{step}} \otimes (\mu^{i,t}-\mbox{best}^t)\otimes randn(1,K)
	\end{equation}
	
	\noindent where $\otimes$ means the entry-wise product of two vectors. The $randn(1,K)$ returns $K$ random numbers from a normal distribution with mean 0 and variance 1. 
	
	Here $\alpha>0$ is the step size scaling factor, which should be related to the scales of the problem of interest.
	
	$\mbox{step}$ is used in the application of L\'evy flights, see in \cite{yang2010nature} for more detail. For standard random walks, use $\mbox{step}=1$.
\item After generating new cuckoos, we update nests $\mu^{i,t}$ as follows:
\begin{equation}\label{new_s}
\mu^{i,t}:=
\begin{cases}
c^{i,t} & \mbox{if } \Psi(c^{i,t}) \leq \Psi(\mu^{i,t})\\
\mu^{i,t} & \mbox{otherwise} 
\end{cases}
\end{equation}

\item A fraction of worse nests are discovered with a probability $p_a$. For that,
we construct new solutions $\mu^{i,t+1}$ as follows (biased/selective random walks):

\begin{equation}\label{new_s_}
\mu^{i,t+1}:=\mu^{i,t}+ [rand()] \otimes [H(p_a-rand())]\otimes (\mu^{a,t}-\mu^{b,t})
\end{equation}

\noindent where  $\mu^{a,t}$ and $\mu^{b,t}$ are two different solutions selected randomly by random permutation. $H$ is Heaviside function. $rand()$ returns a single uniformly distributed random number in the interval (0,1). $[\  ]$ is a $K$ vector.

\end{enumerate}

Each variant of CS manages in a specific way $p_a$, $\alpha$ and $\mbox{step}$.
 
The HMRF-CS combination is summarized in Algorithm \ref{algo}.
\begin{algorithm}\label{algo}
\SetAlgoLined
 Objective function $\Psi(\mu), \mu=(\mu_{1}, ..., \mu_{K})$
 
 Generate initial population of $n$ host nests $\mu^{i,0}$
 
 \While{(t< MaxGeneration) }{
    Compute the best solution using  \ref{best}
    
    Get new cuckoos using  \ref{cuckoos}
    
    Update nests using  \ref{new_s} 
   
    A fraction ($p_a$) of worse nests are abandoned
    
    New nests/solutions are built/generated using \ref{new_s_}
    
    Keep best solutions (or nests with quality solutions)
    
    Update $t:=t+1$
}
 \caption{HMRF-CS algorithm}
\end{algorithm}

\section{Experimental Results}
To show the effectiveness of HMRF-CS method, we have implemented three variants of cuckoo search algorithm: Standard CS \cite{yang2009cuckoo,yang2014cuckoo}, Improved CS \cite{valian2011improved} and Auto Adaptative CS \cite{li2015modified}.

To perform a meaningful study, we have used IBSR and BrainWeb \cite{cocosco1997brainweb} databases where ground truth segmentation is known. The Dice coefficient is used as a statistical validation metric to evaluate the performance.

BrainWeb images are simulated MRI volumes for normal brain from McGill University. These simulations are based on an anatomical model of normal brain. In this database, an image can be selected by setting modality, slice thickness, noise and intensity non-uniformity.

The Internet Brain Segmentation Repository (IBSR) provides manually-guided expert segmented brain data.
	\subsection{Dice Coefficient metric} \label{DC} 
\noindent Dice Coefficient (DC) \cite{1945-dice} measures how much the result is close to the ground truth. Let the resulting class be $\hat{A}$ and its ground truth be $A^*$. Dice Coefficient is given by the following formula:
\begin{equation}
\mathit{DC} 	= 	\frac{ 2 |\hat{A} \cap A^*| }{\ |\hat{A} \cup A^*| } 
\end{equation}
\subsection{Tests context}
Algorithms are implemented in MATLAB 2017a on a computer with Intel Core i7 1.8 GHz CPU, 8G RAM and Microsoft Windows 10.

We have used the BrainWeb database with the parameters: Modality= T1, Slice thickness = 1mm, Noise = 0\% and Intensity non-uniformity = 0\%. The slice 85 is chosen to show results. 

For IBSR database 1-24, we selected the slice 18 to show results. 

In a first stage, we tuned HMRF-CS parameters, and we get: the number of available host nests $n=30$, temperature $T=4$ and $MaxGeneration=100$. In a second stage, we segment the slice 85 in BrainWeb and 18 in IBSR with the three variants of cuckoo search.

Each algorithm is executed ten times. Here, we will show just the best results obtained. 

\subsection{Results}
In this section, we will show different results: execution time, Dice coefficient and the visual results.
 
\subsubsection{Execution time}
For each algorithm, we will give the range of execution time. Table \ref{tab-time} shows range of execution time in the case of BrainWeb image. Table \ref{tab-time_ibsr} shows range of execution time in the case of IBSR image. 
\begin{table}[h!]
	\begin{center}
		\caption{\label{tab-time} range of segmentation time - BrainWeb.}
		\begin{tabular}{|p{4cm}|p{2cm}|}
			\hline
			{Methods} & 
			{Time (s)} \\
			\hline
			
			HMRF-CS\_Auto 		&  130 - 140 \\
			\hline
			HMRF-CS\_Standard 	&  85 - 105 \\
			\hline
			HMRF-CS\_Improved & 85 - 105 \\
			\hline
		\end{tabular}
	\end{center}
\end{table}

\begin{table}[h!]
	\begin{center}
		\caption{\label{tab-time_ibsr} range of segmentation time - IBSR.}
		\begin{tabular}{|p{4cm}|p{2cm}|}
			\hline
			{Methods} & 
			{Time (s)} \\
			\hline
			
			HMRF-CS\_Auto 		&  230 -250 \\
			\hline
			HMRF-CS\_Standard 	& 125 - 150 \\
			\hline
			HMRF-CS\_Improved & 125 - 150 \\
			\hline
		\end{tabular}
	\end{center}
\end{table}
\subsubsection{Dice coefficient}
We will show in each case, the three best tests. 
Tables \ref{tab-dc}, \ref{tab-dc_2} and \ref{tab-dc_3} show DC values of the three variants of CS with BrainWeb image. In each column, the best results are given in bold type.

\begin{table}[!h]
	\centering
	\caption{\label{tab-dc} DC values - BrainWeb - test 1.}
	\begin{tabular}{|l|c|c|c|c|}
		\hline
		\multirow{2}{*}{\textbf{  Methods  }}  			& \multicolumn{4}{c|}{\textbf{ Dice Coefficient }} \\
		\cline{2-5}
		& \textbf{ GM } & \textbf{ WM } 		& \textbf{ CSF } 		& \textbf{ Mean } \\
		\hline
        HMRF-CS\_Auto & 0.963 	& 0.988 	& 0.973 	& 0.975 \\
		\hline
		HMRF-CS\_Standard & \textbf{0.968} & 0.988  	 	& 0.972 	& 0.976\\
		\hline
     	HMRF-CS\_Improved 	& 0.967 	& \textbf{0.989} 	& \textbf{0.979} 	& \textbf{0.978} \\
		\hline

	\end{tabular}
\end{table}

\begin{table}[!h]
	\centering
	\caption{\label{tab-dc_2} DC values - BrainWeb - test 2.}
	\begin{tabular}{|l|c|c|c|c|}
		\hline
		\multirow{2}{*}{\textbf{  Methods  }}  			& \multicolumn{4}{c|}{\textbf{ Dice Coefficient }} \\
		\cline{2-5}
		& \textbf{ GM } & \textbf{ WM } 		& \textbf{ CSF } 		& \textbf{ Mean } \\
		\hline
		HMRF-CS\_Auto & 0.961	& 0.985 	& 0.974 	& 0.973 \\
		\hline
		HMRF-CS\_Standard & 0.950 & 0.969  	 	& \textbf{0.979}	& 0.966\\
		\hline
		HMRF-CS\_Improved 	& \textbf{0.963} 	& \textbf{0.986} 	& \textbf{0.979} 	& \textbf{0.976} \\
		\hline
		
	\end{tabular}
\end{table}

\begin{table}[!h]
	\centering
	\caption{\label{tab-dc_3} DC values - BrainWeb - test 3.}
	\begin{tabular}{|l|c|c|c|c|}
		\hline
		\multirow{2}{*}{\textbf{  Methods  }}  			& \multicolumn{4}{c|}{\textbf{ Dice Coefficient }} \\
		\cline{2-5}
		& \textbf{ GM } & \textbf{ WM } 		& \textbf{ CSF } 		& \textbf{ Mean } \\
		\hline
		HMRF-CS\_Auto & 0.939 	& 0.974 	& 0.944 	& 0.952 \\
		\hline
		HMRF-CS\_Standard & 0.943 & 0.966  	 	& \textbf{0.977} 	& 0.962\\
		\hline
		HMRF-CS\_Improved 	& \textbf{0.961} 	& \textbf{0.985} 	& 0.974 	& \textbf{0.974} \\
		\hline
		
	\end{tabular}
\end{table}

Tables \ref{tab-dc_ibsr}, \ref{tab-dc_ibsr_2} and \ref{tab-dc_ibsr_3} show DC values of the three variants of CS with IBSR image.

\begin{table}[!h]
	\centering
	\caption{\label{tab-dc_ibsr} DC values - IBSR - test 1.}
	\begin{tabular}{|l|c|c|c|c|}
		\hline
		\multirow{2}{*}{\textbf{  Methods  }}  			& \multicolumn{4}{c|}{\textbf{ Dice Coefficient }} \\
		\cline{2-5}
		& \textbf{ GM } & \textbf{ WM } 		& \textbf{ CSF } 		& \textbf{ Mean } \\
		\hline
		HMRF-CS\_Auto & \textbf{0.932} 	& 0.933 	& 0.544 	& 0.803 \\
		\hline
		HMRF-CS\_Standard & 0.927 & \textbf{0.934} 	 	& \textbf{0.552} 	& \textbf{0.805}\\
		\hline
		HMRF-CS\_Improved 	& 0.924 	& \textbf{0.934} 	& 0.546 	& 0.801 \\
		\hline
		
	\end{tabular}
\end{table}
\begin{table}[!h]
	\centering
	\caption{\label{tab-dc_ibsr_2} DC values - IBSR - test 2.}
	\begin{tabular}{|l|c|c|c|c|}
		\hline
		\multirow{2}{*}{\textbf{  Methods  }}  			& \multicolumn{4}{c|}{\textbf{ Dice Coefficient }} \\
		\cline{2-5}
		& \textbf{ GM } & \textbf{ WM } 		& \textbf{ CSF } 		& \textbf{ Mean } \\
		\hline
		HMRF-CS\_Auto & \textbf{0.927} 	& \textbf{0.934} 	& \textbf{0.547} 	& \textbf{0.802} \\
		\hline
		HMRF-CS\_Standard & 0.920 & 0.933  	 	& 0.544 	& 0.799\\
		\hline
		HMRF-CS\_Improved 	& 0.923 	& \textbf{0.934} 	& 0.544 	& 0.800 \\
		\hline
		
	\end{tabular}
\end{table}
\begin{table}[!h]
	\centering
	\caption{\label{tab-dc_ibsr_3} DC values - IBSR - test 3.}
	\begin{tabular}{|l|c|c|c|c|}
		\hline
		\multirow{2}{*}{\textbf{  Methods  }}  			& \multicolumn{4}{c|}{\textbf{ Dice Coefficient }} \\
		\cline{2-5}
		& \textbf{ GM } & \textbf{ WM } 		& \textbf{ CSF } 		& \textbf{ Mean } \\
		\hline
		HMRF-CS\_Auto & 0.909 	& 0.917 	& \textbf{0.552} 	& 0.793 \\
		\hline
		HMRF-CS\_Standard & 0.915 & 0.928  	 	& 0.544 	& 0.796\\
		\hline
		HMRF-CS\_Improved 	& \textbf{0.921} 	& \textbf{0.933} 	& 0.544 	& \textbf{0.799} \\
		\hline
		
	\end{tabular}
\end{table}

\subsubsection{The visual results}
\begin{figure}[!h]
	\renewcommand{\arraystretch}{1}
	
	\setlength{\tabcolsep}{2pt}
	\setlength{\extrarowheight}{0pt}%
	
	\footnotesize
	\newlength{\iw}
	\newlength{\ih}
	\renewcommand{\angle}{90}
	\setlength{\iw}{2.9cm}
	\setlength{\ih}{6cm}
	\centering
	\begin{tabular}{|>{\centering\arraybackslash}m{1cm}|>{\centering\arraybackslash}m{6cm}|}
		\hline
		(a) &
		\includegraphics[width=\iw,height=\ih,angle=\angle]{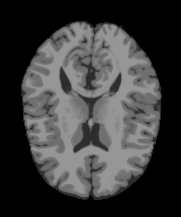}  
		\\
		
		\hline
		(b) &
		\includegraphics[width=\iw,height=\ih,angle=\angle]{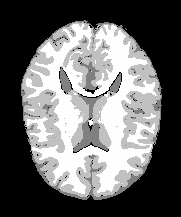}  
		\\
		\hline
		(b) &
		\includegraphics[width=\iw,height=\ih,angle=\angle]{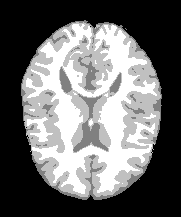}  
		\\
		\hline
		
	\end{tabular}
	
	\caption{\textbf{(a)} image to segment fom BrainWeb database, \textbf{(b)} ground truth image, \textbf{(c)} segmented image with HMRF-CS\_Improved.}
	\label{fig-brainweb}
	
\end{figure}

\begin{figure}[!h]
	\renewcommand{\arraystretch}{1}

		\setlength{\tabcolsep}{2pt}
		\setlength{\extrarowheight}{0pt}%
		
		\footnotesize
		\renewcommand{\angle}{90}
		\setlength{\iw}{2.9cm}
		\setlength{\ih}{6cm}
		\centering
		
		\begin{tabular}{|>{\centering\arraybackslash}m{1cm}|>{\centering\arraybackslash}m{6cm}|}
			\hline
			(a) &
			\includegraphics[width=\iw,height=\ih,angle=\angle]{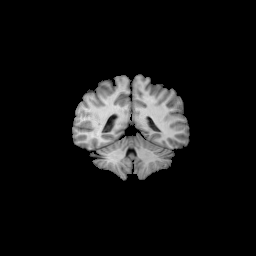}  
			\\

			\hline
			(b) &
			\includegraphics[width=\iw,height=\ih,angle=\angle]{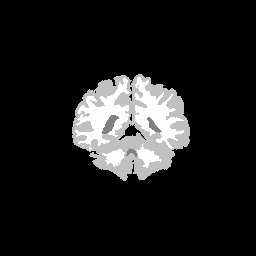}  
		\\
			\hline
		(b) &
		\includegraphics[width=\iw,height=\ih,angle=\angle]{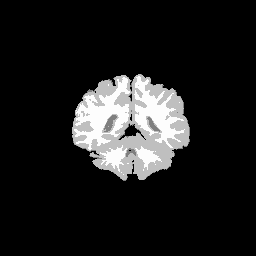}  
		\\
			\hline
			
		\end{tabular}
		
		\caption{\textbf{(a)} image to segment from IBSR database, \textbf{(b)} ground truth image, \textbf{(c)} segmented image using HMRF-CS\_Improved.}
		\label{fig-ibsr}

\end{figure}

In this section, we will show the visual results using HMRF-CS\_Improved method. Figure \ref{fig-brainweb} shows BrainWeb image to segment, ground truth image and the segmented image. Figure \ref{fig-ibsr} shows IBSR image to segment, ground truth image and the segmented image. 
 
\section{CONCLUSIONS}

In this paper, we have presented a new approach, that combines Hidden Markov Random Fields and cuckoo search to perform segmentation on MRI images. 
The tests conducted have focused on the brain images from largely used databases: BrainWeb and IBSR.
The HMRF-CS\_Improved combination outperforms  other combination  methods  tested that are: HMRF-CS\_Standard and CS\_Auto. CS is simple, fast, robust and accurate.
However, the proposed method needs to be tested on other images. A comparative study with other state-of-the-art segmentation methods also has to be conducted.  A statistical study of parameters: $n, T, \mbox{step}, \alpha, p_a$ is needed.

\bibliographystyle{IEEEbib}
\bibliography{strings}

\end{document}